\documentclass[aps,prl,twocolumn,amsfonts,amsmath,amssymb,superscriptadd
ress]{revtex4-1}%
\pdfoutput=1

\usepackage[colorlinks]{hyperref}
\usepackage{amsfonts}
\usepackage{color}
\usepackage{amsmath}
\usepackage{amssymb}
\usepackage{graphicx}%
\usepackage{soul}

\begin{document}
\title{
Fundamental precision limit of a Mach-Zehnder interferometric sensor when one of the inputs is the vacuum}

%\date{\today}
%
\author{Masahiro Takeoka}

\affiliation{National Institute of Information and Communications Technology,
Koganei, Tokyo 184-8795, Japan}

\author{Kaushik P. Seshadreesan}

\affiliation{Max Planck Institute for the Science of Light, 91058 Erlangen, Germany}

\affiliation{Hearne Institute for Theoretical Physics and Department of Physics
and Astronomy, Louisiana State University, Baton Rouge, Louisiana
70803, USA}

\affiliation{National Institute of Information and Communications Technology,
Koganei, Tokyo 184-8795, Japan}

\author{Chenglong You}

\affiliation{Hearne Institute for Theoretical Physics and Department of Physics
and Astronomy, Louisiana State University, Baton Rouge, Louisiana
70803, USA}

\author{Shuro Izumi}

\affiliation{National Institute of Information and Communications Technology,
Koganei, Tokyo 184-8795, Japan}

\affiliation{Sophia University, 7-1 Kioicho, Chiyoda-ku, Tokyo 102-8554, Japan}

\author{Jonathan P. Dowling}

\affiliation{Hearne Institute for Theoretical Physics and Department of Physics
and Astronomy, Louisiana State University, Baton Rouge, Louisiana
70803, USA}

%\pacs{42.50.Dv, 42.50.Ex, 03.67.-a, 03.65.Wj}

%42.50.Dv Quantum state engineering and measurements 
%42.50.Ex Optical implementations of quantum information processing 
%and transfer 
%03.67.-a Quantum information  
%03.65.Wj State reconstruction, quantum tomography  

\begin{abstract}
In the lore of quantum metrology, one often hears (or reads) the following no-go theorem: 
If you put vacuum into one input port of a balanced Mach-Zehnder Interferometer, then no matter 
what you put into the other input port, and no matter what your detection scheme, the sensitivity can never be better than the 
shot noise limit (SNL). Often the proof of this theorem is cited to be in Ref. 
[C. Caves, Phys. Rev. D 23, 1693 (1981)], but upon further inspection, no such claim is made there. A 
quantum-Fisher-information-based argument suggestive of this no-go theorem appears in Ref. [M. Lang and C. Caves,
Phys. Rev. Lett. 111, 173601 (2013)], but is not stated in its full generality. Here we thoroughly explore this no-go theorem and give the rigorous statement: the no-go theorem holds whenever the unknown phase shift is split between both arms of the interferometer, but remarkably does not hold when only one arm has the unknown phase shift. In the latter scenario, we provide an explicit measurement strategy that beats the SNL. We also point out that these two scenarios are physically different and correspond to different types of sensing applications.
\end{abstract}

\maketitle

\begin{figure}
\begin{center}
\includegraphics[width=80mm]{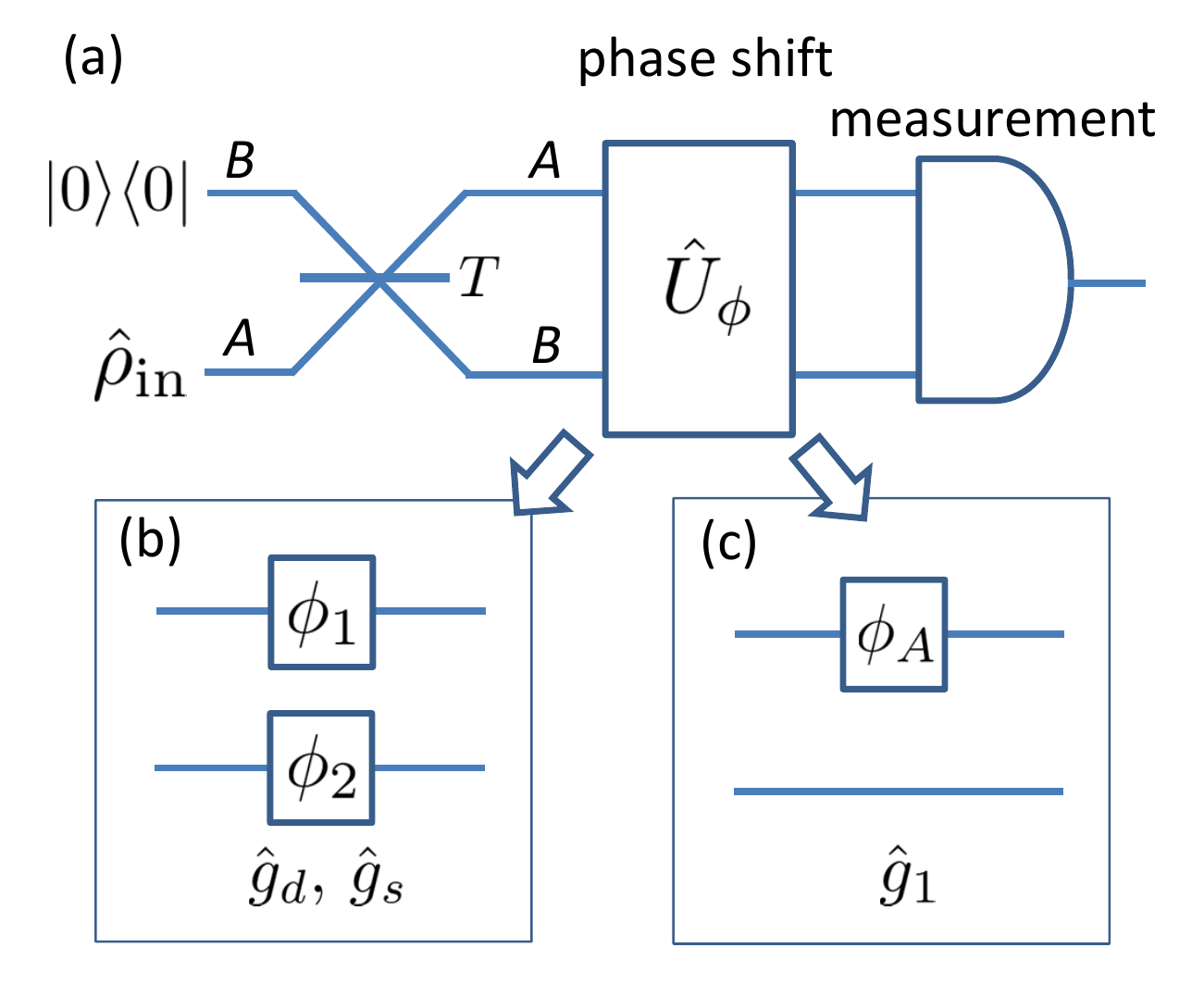}   %
\caption{\label{fig:mz}
(a) Mach-Zehnder interferometer phase estimation and 
the two different phase shift models: 
phase shift(s) are applied in (b) two arms, or 
(c) one arm of the interferometer. 
See the main text for the details. 
}
\end{center}
\end{figure}

{\it Introduction.---} 
In the field of quantum metrology\cite{GLM11,DJK15,DS15}, 
a Mach-Zehnder interferometer (MZI) is a tried and true 
workhorse that has the additional advantage that any result obtained for it also applies to a Michelson interferometer (MI) and hence has a potential application to gravitational wave detection. In most current implementations of gravitational wave detectors, the MI is fed with a strong coherent state of light in one input port and vacuum in the other (Fig. \ref{fig:mz}).  It was in this context that Caves in 1981 \cite{caves81} showed that such a design would always only ever achieve the shotnoise limit (SNL). 
Then he showed if you put squeezed vacuum into the unused port, you could beat the SNL. Several implementations of this squeezed vacuum scheme have already been demonstrated in the GEO 600 gravitational detector, and plans are underway to utilize this approach in the LIGO and VIRGO detectors in the future \cite{GEO600,GEO6001}.

It then appeared, that in the lore of quantum metrology, this result was extended --- without proof --- to the following no-go theorem: If you put quantum vacuum into one input port of a balanced MZI, then no matter what quantum state of light you put into the other input port, and no matter what your detection scheme, the sensitivity can never be better than the SNL. Often the proof of this theorem is cited to be the original 1981 paper by Caves \cite{caves81}, but upon further inspection, no such general claim is made 
there. A quantum-Fisher-information-based argument suggestive of this no-go theorem appeared in Ref. \cite{LC13} by Lang and Caves, but it does not explore the statement in adequate generality.

In this work, we give a full statement of the no-go theorem. The statement proved here is the following: if the unknown phase shifts are in both of the two arms of the MZI, then the no-go theorem holds no matter whether the MZI is balanced or not. However, in the case where the unknown phase shift is in only one arm of the MZI, then the no-go theorem does not necessarily hold. The former is a multiparameter measurement and the latter is a single parameter. For the latter, we show an explicit scheme with a probe and measurement that can beat the SNL in the sense that its classical Fisher information (CFI) is proportional to the square of the total photon number used at the input and the measurement. The underlying issue is that two different models for the unknown phase shift unitary operation in the MZI can give different values of the QFI \cite{JD12,PHS15}. Since only the phase difference is utilized in both models, it has been thought that this discrepancy is a flaw in the interpretation of the QFI \cite{JD12} or is related to the assumptions of the input states and the measurements \cite{PHS15}. By contrast, here we point out that the different unitaries correspond to physically different types of sensors, and their choice should depend on the concrete application scenarios. 
Also for the former scenario (i.e. unknown phase shifts in two arms), we show that one has to carefully consider 
the phase sum (often regarded as the ``global phase'' though) whereas only phase difference is the quantity of interest. 
In other words, it is intrinsically a two-parameter estimation problem.  

Related to the above, we also point out the pitfalls of using only the quantum Fisher information (QFI), or the closely related quantum Cram\'{e}r-Rao (QCRB) bound \cite{HelstromBook}, to make claims of a quantum metrological advantage, without explicitly providing a detection scheme that would actually achieve that advantage \cite{JD12}. Before the QFI approach came into vogue in recent years, often theorists would try to optimize the input state and the detection scheme simultaneously. This often led to input states and detection schemes difficult to implement. The QFI approach freed us from having to optimized over all detection schemes, more accurately over all Positive Operate Valued Measures (POVM), but that freedom, carried a very high cost. The issue is that the optimal POVM that achieves the QFI may be difficult to implement or contain hidden resources, such as a strong local oscillator, that are not fairly counted as far as a quantum advantage is concerned \cite{JD12}.

%%%%%%%%%%%%%%%%%%%%%%%%%%%%%%%%%%%%%%%%%%%
%
%%%%%%%%%%%%%%%%%%%%%%%%%%%%%%%%%%%%%%%%%%%
%

{\it Quantum Fisher information approach to phase sensing---}
A schematic of the Mach-Zehnder (MZ) interferometer-type sensing 
we consider here is illustrated in Fig.~\ref{fig:mz}(a). 
Two input modes A and B are interfered via a beam splitter 
with transmittance $T$, and then put into the phase shift unitary 
operation $\hat{U}_\phi$ followed by some measurement. 
In addition to this standard setting, we restrict one of the input states 
to always be the quantum vacuum state, whereas the other input can be an arbitrary quantum state (possibly mixed). 

A similar setup to the one we consider here has recently used 
by Lang and Caves \cite{LC13}, where the two input ports 
are assumed to be in an arbitrary pure state $|\chi\rangle$ 
and a coherent state $|\alpha\rangle$, and the beam splitter transmittance 
is chosen to be 50/50 (i.e. $T=1/2$). The phase shift unitary operator is 
$\hat{U}_\phi = e^{i \hat{g}_s \phi_s} e^{i \hat{g}_d \phi_d}$, 
where $\phi_s$ and $\phi_d$ are the phase sum and difference 
of the two modes, respectively, 
$\hat{g}_s = (\hat{a}^\dagger \hat{a} + \hat{b}^\dagger \hat{b})/2$,  
$\hat{g}_d = (\hat{a}^\dagger \hat{a} - \hat{b}^\dagger \hat{b})/2$. 
These two phase shift parameters reflect the unknown phase shifts in 
the two arms of the MZI, $\phi_1$ and $\phi_2$, 
as $\phi_s = \phi_1 + \phi_2$, $\phi_d = \phi_1 - \phi_2$ 
(see Fig.~\ref{fig:mz}(b)). 
$\hat{a}^\dagger$ ($\hat{b}^\dagger$) and 
$\hat{a}$ ($\hat{b}$) are creation and annihilation operators in 
mode $A$ ($B$), respectively. 

Then the authors showed that for a coherent state input 
with $\alpha = 0$, i.e. for the vacuum input, 
the quantum Fisher information (QFI) for the phase 
difference turns out to be the average photon number of the input: 
%%%%%%%%%%%%%%%%%%%%%%%%%%%
\begin{equation}
\label{eq:qfi_LC13}
F_Q(|\chi\rangle, \hat{g}_d) 
= \langle\chi| \hat{n} |\chi\rangle = \bar{n}_\chi , 
\end{equation}
%%%%%%%%%%%%%%%%%%%%%%%%%%%
where $\hat{n}=\hat{a}^\dagger \hat{a}$. 
This result suggests that the precision of 
the phase sensing is shot-noise limited, 
when one of the input ports contains only vacuum 
(and the other mode contains any pure state), since the QCRB is $\Delta^2\phi\geq1/F_Q$.

However, the above result does not answer questions such as whether the no-go theorem still holds when the interferometer is not balanced, or when the phase shift unitary operator is chosen differently. Firstly, for the phase shift unitary operator $\hat{g}_d$, when $T$ deviates from $1/2$, the QCRB already appears to beat the SNL. Keeping $T$ as a free parameter, and using the fact that the QFI of a pure state in estimating a phase shift generated by a generator $\hat{g}$ is given by 
$4\left(\langle \hat{g}^{2}\rangle -\langle \hat{g}\rangle ^{2}\right)$, we arrive at
%%%%%%%%%%%%%%%%%%%%%%%%%%%
\begin{equation}
\label{eq:qfi_g_d}
F_Q(|\chi\rangle, \hat{g}_d, T) 
= \{1-(1-2T)^2\} \bar{n}_\chi + (1-2T)^2 V_\chi.
\end{equation}
%%%%%%%%%%%%%%%%%%%%%%%%%%%	
(See Supplemental Material 1 for the derivation.) This beats the SNL for any non-50/50 beam splitter quite spectacularly. For example, with $T\rightarrow 0$, the QCRB 
approaches $\Delta^2\phi=1/V_x<1/\bar{n}_\chi$ for some inputs such as squeezed vacuum \cite{ARCPHLD10}.

Secondly, as pointed out and rigorously discussed in Ref. \cite{JD12}, a different choice of the phase shift unitary can give a different 
value for the QFI. 
For example, in lieu of the phase shift operator $\hat{g}_d$, one can instead choose $\hat{U}_\phi = e^{ i \hat{g}_1 \phi_A}$, 
where $\hat{g}_1 = \hat{a}^\dagger \hat{a}$, such that phase shift is generated only in one arm. The QFI for the phase shift unitary operator $\hat{g}_1$ is found to be
%%%%%%%%%%%%%%%%%%%%%%%%%%%
\begin{equation}
\label{eq:qfi_g_1}
F_Q(|\chi\rangle, \hat{g}_1) 
= \bar{n}_\chi + V_\chi ,
\end{equation}
%%%%%%%%%%%%%%%%%%%%%%%%%%%
where $V_\chi = \langle\chi| \hat{n}^2 |\chi\rangle 
- \langle\chi| \hat{n} |\chi\rangle^2$ is the photon number 
variance of $|\chi\rangle$. (See Supplemental Material 1 for the derivation.) 
This is obviously different from Eq.~(\ref{eq:qfi_LC13}), and again 
implies a sub-SNL result, since $V_\chi > \bar{n}_\chi$ 
is possible for some inputs, as mentioned above.

These results extrapolated from Ref. \cite{LC13} are thus perplexing, since 
seemingly, both Eqs.~(\ref{eq:qfi_g_d}) and 
(\ref{eq:qfi_g_1}) suggest the possibility of 
sub-SNL precision phase sensing even with the vacuum input 
 into one of the input ports.

{\it Phase shift in both arms vs. in one arm in the MZI sensing---}
We point out that the above phase-shift unitary operators 
(Figs.~\ref{fig:mz}(b) and (c)) 
have different physical meanings and their choice 
should depend on what type of application scenario is in your mind. 
For the gravitational wave detection application, $\hat{g}_d$ and 
$\hat{g}_s$ should be chosen since the two arms of the (Michelson) 
interferometer both have unknown phase shifts induced by the gravitational waves 
(Fig.~\ref{fig:mz}(b)). Also some commonly used sensing devices such as 
a differential interference contrast microscope \cite{N55} 
should be modeled in the same way. (See also its quantum version \cite{OOT13}.)

On the other hand, the most primitive use of the Mach-Zehnder interferometer 
is to put a sample in one of the two arms to measure the corresponding phase shift. 
This configuration is also widely used as a simple and low-cost technology 
to measure the sample's density distribution, pressure, temperature, etc. 
This type of sensor should be modeled by $\hat{g}_1$ (Fig.~\ref{fig:mz}(c)). 
Since these two models are physically different, they may lead to 
different outcomes in our problem; the MZI with vacuum in one input 
port. That is, they could have different fundamental precision limits 
with vacuum in one input port. 
We will rigorously analyze each model in the following. 

{\it Remedy: Full quantum Fisher information matrix treatment---}
The MZI sensing with the $\hat{g}_s$-$\hat{g}_d$ model in its full generality is a two-parameter estimation problem 
since there are two unknown parameters, $\phi_s$ and $\phi_d$, in the system 
(although usually only the phase difference $\phi_d$ is an interesting quantity to measure). 
Therefore, a two-by-two quantum Fisher information matrix (QFIM) is considered. 
The problem in Eq. (\ref{eq:qfi_g_d}) is in fact due to the ignorance of the phase sum
$\phi_s$ \footnote{In Ref.~\cite{LC13}, the QFIM of the system considered was calculated. 
However, they reduce it to the single-parameter estimation 
(i.e. drop off the terms for $\phi_s$) which looses 
the tightness of the bound. Note that this problem does not appear 
in Eq.~(\ref{eq:qfi_LC13}) since with $T=1/2$, the non-diagonal term of 
the QFIM goes to zero and thus the problem reduces to two independent 
single-parameter estimations. Nevertheless, in Ref.~\cite{LC13}, 
they also consider the non-vacuum input case where 
the bound may have some looseness.
}.
In multi-parameter estimation, the QCRB is given by 
%%%%%%%%%%%%%%%%%%%%%%%%%%%
\begin{equation}
\label{eq:multi_QCRB}
\Sigma \ge \frac{\mathcal{F}_Q^{-1}}{m}, 
\end{equation}
%%%%%%%%%%%%%%%%%%%%%%%%%%%
where $\Sigma$ is the covariance matrix of the estimator 
including both $\phi_s$ and $\phi_d$, $m$ is the number of trials, 
and $\mathcal{F}_Q$ is the two-by-two QFIM: 
%%%%%%%%%%%%%%%%%%%%%%%%%%%
\begin{equation}
\label{eq:QFIM}
\mathcal{F}_Q = \left[
\begin{array}{cc}
F_{dd} & F_{sd} \\
F_{ds} & F_{ss} 
\end{array}
\right] ,
\end{equation}
%%%%%%%%%%%%%%%%%%%%%%%%%%%
where $s$ and $d$ correspond to $\phi_s$ and $\phi_d$. 
The first diagonal element of $\mathcal{F}_Q^{-1}$ 
in Eq. (\ref{eq:multi_QCRB}) 
corresponds to the estimation limit of $\phi_d$, which is explicitly given by 
%%%%%%%%%%%%%%%%%%%%%%%%%%%
\begin{equation}
\label{eq:QFIM_dd}
\frac{F_{ss}}{F_{ss} F_{dd} - F_{sd} F_{ds}} .
\end{equation}
%%%%%%%%%%%%%%%%%%%%%%%%%%%

For an arbitrary mixed quantum state, the QFIM is in general not easy to calculate. 
However, the optimal input state that maximizes $\mathcal{F}_Q$ 
is always given by a pure-state input. 
This is the consequence of the convexity of the QFIM: 
for $\hat{\rho}_\phi = p \hat{\sigma}_\phi + (1-p) \hat{\tau}_\phi$, 
%%%%%%%%%%%%%%%%%%%%%%%%%%%
\begin{equation}
\label{eq:convexity}
\mathcal{F}_Q(\hat{\rho}_\phi) \le p \mathcal{F}_Q(\hat{\sigma}_\phi) 
+ (1-p) \mathcal{F}_Q(\hat{\tau}_\phi),  
\end{equation}
%%%%%%%%%%%%%%%%%%%%%%%%%%%
holds. This can be proved by using the monotonicity of the QFIM under 
the completely positive trace preserving (CPTP) map \cite{P96,PG10} 
and extending the proof of the convexity for the QFI \cite{F01}. 
(See Supplementary Material 2.) 
The statement basically says that a statistical mixture of the input states 
will never increase the QFIM and thus implies that the QFIM is maximized 
with a pure state input. 
The optimal pure state for the QFIM is also optimal for 
the multi-parameter QCRB (\ref{eq:multi_QCRB}) 
since the QFIM is a positive matrix 
and for positive matrices $A$ and $B$, 
$B^{-1} \ge A^{-1}$ holds if and only if $A \ge B$. 
That is, a statistical mixture of the input states 
will never increase the sensitivity for both single- and 
multiple-parameter estimation.

Therefore, by considering a pure input state $|\chi\rangle$, the elements of the QFIM are given by 
%%%%%%%%%%%%%%%%%%%%%%%%%%%
\begin{equation}
\label{eq:QFIM_entry}
F_{ij} = 4 \left( \langle \hat{g}_i \hat{g}_j \rangle 
- \langle \hat{g}_i \rangle \langle \hat{g}_j \rangle \right), 
\end{equation}
%%%%%%%%%%%%%%%%%%%%%%%%%%%
where $i,j$ takes $s$ and $d$. 
These are explicitly given by 
%%%%%%%%%%%%%%%%%%%%%%%%%%%
\begin{eqnarray}
\label{eq:QFIM_entry_dd}
F_{dd} & = & \{1-(1-2T)^2\} \bar{n}_\chi + (1-2T)^2 V_\chi, 
\\
\label{eq:QFIM_entry_ss}
F_{ss} & = & V_\chi, 
\\
\label{eq:QFIM_entry_ds}
F_{ds} & = & F_{sd} = -(1-2T) V_\chi, 
\end{eqnarray}
%%%%%%%%%%%%%%%%%%%%%%%%%%%
where note that $F_{dd}$ corresponds to Eq.~(\ref{eq:qfi_g_d}). 
Inserting these into (\ref{eq:QFIM_dd}), we get 
%%%%%%%%%%%%%%%%%%%%%%%%%%%
\begin{equation}
\label{eq:no-go_g_d}
\Delta^2 \phi_d \ge \frac{1}{4T(1-T)m\bar{n}_\chi}, 
\end{equation}
%%%%%%%%%%%%%%%%%%%%%%%%%%%
where the minimum of the right hand side is obtained with $T=1/2$
as $1/(m \bar{n}_\chi)$, which is the SNL, as it should be. 
That is, no matter how highly nonclassical the input
state $\hat{\rho}_{\rm in}$ is, and no matter what POVM you deploy, the SNL cannot be surpasses 
for $\hat{g}_d$ so long as the other input to the interferometer 
is the vacuum state. Thus, this result establishes the no-go theorem in a most general form, 
which includes the beam splitter transmissivity as a free parameter.

{\it Phase shift in one arm ($\hat{g}_1$)---}
The $\hat{g}_1$ model is a single-parameter estimation problem and 
thus (\ref{eq:qfi_g_1}) is directly applied to the QCRB, which 
suggests the sub-SNL sensitivity with high $V_\chi$, that is, 
input states with high photon number fluctuation such as squeezed vacuum. 
Then as mentioned at the introduction, the QFI-only approach may have 
the pitfall that the optimal POVM attaining the QCRB could contain huge amount 
of hidden resources as pointed out by Jarzyna and Demkowicz-Dobrza{\'n}ski 
\cite{JD12}. 
In other words one can fool oneself into thinking, 
via the QFI-only approach, that there is some 
quantum metrological advantage, where none actually exists.

There are two remedies. 
The first, and the one we recommend, is that if authors wish to claim 
a quantum metrological advantage from a QFI-only calculation, 
they then must provide a detection scheme that actually hits 
the related QCRB, so all resources hidden in the associated POVM may be 
then laid bare for all to see. (We should note that in Ref.~\cite{ARCPHLD10}, the authors 
were careful to back up the QFI calculation by providing 
a detection scheme --- the parity operator --- that actually hits the QCRB.)

The second remedy, besides producing the POVM that hits the limit, 
is to rule out any external resource that might give some phase information to 
the measurement device. Such a ``rule-out" protocol 
was introduced by Jarzyna and Demkowicz-Dobrza{\'n}ski \cite{JD12}. 
They resolved this
issue by introducing the idea of a phase-averaged input state, where 
the two-mode input state from the two input ports is 
averaged by a common phase shift, which preserves the relative phase 
between two modes, but does not allow any phase information to be brought in
from the outside of the interferometer, e.g. 
from the measurement devices themselves (a similar discussion appeared 
in the context of superselection rule \cite{PHS15}). 
Therefore, the QFI of the phase-averaged input gives  
the proper phase-sensing limit without any external phase reference. 
A simple way to understand the phase averaging is to think of it 
as a type of phase randomization akin to preparing a thermal state. 
A thermal state can be used in an MZI for SNL interferometry, even 
though it contains no coherence, because each photon 
--- as in Dirac's dictum --- 
only ever interferes with itself. 
In this way the advantage of any hidden resource in the POVM is mitigated.

Here we apply these two remedies separately. 
First, we employ the phase-averaging approach to eliminate any hidden resource 
in the POVM. 
For the two input states, we consider a vacuum and 
an arbitrary quantum state with the density matrix of 
%%%%%%%%%%%%%%%%%%%%%%%%%%%
\begin{equation}
\label{eq:input_state}
\hat{\rho}_{\rm in} = \sum_{n,m=0}^\infty c_{nm} |n \rangle\langle m|, 
\end{equation}
%%%%%%%%%%%%%%%%%%%%%%%%%%%
where $|n\rangle$ is the $n$-photon number state. 
Then the phase-averaged input is given by
%%%%%%%%%%%%%%%%%%%%%%%%%%%
\begin{eqnarray}
\label{eq:phase_averaged_input}
\Psi_{\textrm{avg}} & = & 
\int\frac{d\theta}{2\pi} \hat{V}^A_\theta \hat{V}^B_\theta 
\left( \hat{\rho}_{\rm in}^A \otimes |0 \rangle\langle 0|^B \right) 
\hat{V}^{A \, \dagger}_\theta \hat{V}^{B \, \dagger}_\theta 
\nonumber\\ & = & 
\sum_{n,m=0}^{\infty}\int\frac{d\theta}{2\pi}
e^{i\theta (n-m)} c_{nm} |n\rangle \langle m|^A 
\otimes |0\rangle \langle 0|^B 
\nonumber\\ & = & 
\sum_{n=0}^{\infty} p_n
|n\rangle \langle n|^A \otimes |0\rangle \langle 0|^B ,
\end{eqnarray}
%%%%%%%%%%%%%%%%%%%%%%%%%%%
where $\hat{V}^A_\theta = e^{i\theta \hat{a}^\dagger \hat{a}}$, 
$\hat{V}^B_\theta = e^{i\theta \hat{b}^\dagger \hat{b}}$, and 
$p_n = c_{nn}$ is a real positive number satisfying 
$\sum_n p_n =1$. 

The state after the first beamsplitter of the MZI and the phase shifting 
is given by 
%%%%%%%%%%%%%%%%%%%%%%%%%%%
\begin{eqnarray}
\label{eq:phase_averaged_output}
\Psi_{\textrm{avg}}^\phi & = & 
\hat{U}_\phi^{(1)\, AB} \hat{B}^{AB}_T \Psi_{\textrm{avg}}
\hat{B}_T^{\dagger AB} \hat{U}_\phi^{(1)\dagger \, AB}
\nonumber\\ & = & 
\sum_{n=0}^\infty 
p_n |\psi_n(\phi)\rangle\langle\psi_n(\phi)|_{AB} ,
\end{eqnarray}
%%%%%%%%%%%%%%%%%%%%%%%%%%%
where 
%%%%%%%%%%%%%%%%%%%%%%%%%%%
\begin{eqnarray}
|\psi_n(\phi)\rangle _{AB} & = & 
\sum_{j=0}^{n}e^{-ij\phi}\sqrt{{n \choose j}}
\nonumber\\ && \times 
\sqrt{T}^{j}\sqrt{1-T}^{n-j}|j\rangle _A\otimes|n-j\rangle_B. 
\end{eqnarray}
%%%%%%%%%%%%%%%%%%%%%%%%%%%

By using the convexity of the QFI and noticing that 
$|\psi_n(\phi)\rangle$ and $|\psi_{n'}(\phi)\rangle$ 
are orthogonal for $n \ne n'$, we have 
%%%%%%%%%%%%%%%%%%%%%%%%%%%
\begin{equation}
F^{(1)}_{Q}\left(\Psi_{\textrm{avg}}^\phi\right)
= \sum_{n=0}^{\infty} p_n 
F^{(1)}_{Q}\left(|\psi_n(\phi)\rangle \right) ,
\end{equation}
%%%%%%%%%%%%%%%%%%%%%%%%%%%
where 
\begin{eqnarray}
\label{eq:F_Q_n}
F_{Q}^{\left(1\right)} \left(|\psi_n(\phi)\rangle\right)
& = & 4\left(\langle \hat{g}_1^2\rangle -\langle \hat{g}_{1}\rangle 
^{2}\right)
\nonumber\\ & = & 4\bar{n}T\left(1-T\right).
\end{eqnarray}
(See Supplementary Material 3 for the detailed derivation.) 
The maximum in Eq. (\ref{eq:F_Q_n}) is attained at $T=1/2$, 
and is equal to $\bar{n}$, as it should be.

Consequently, the QFI for $\Psi^\phi_{\rm ave}$ is given as 
\begin{equation}
F_{Q}\left(\Psi_{\textrm{{avg}}}^\phi\right) = 
\sum_{n=0}^{\infty} n p_n = \bar{n} , 
\end{equation}
where $\bar{n}$ is the average photon number of $\hat{\rho}_{\rm in}$, 
and thus we find that the phase sensitivity is lower bounded as
\begin{equation}
\Delta^2 \phi_A\geq\frac{1}{m \bar{n}}.
\end{equation}
That is, if the optimal POVM is not allowed to have external phase information,
 the estimation precision is limited by the shot-noise limit. 

There is, however, one question remaining: if one is allowed to use 
some additional resource at the measurement, is it possible to surpass the SNL 
with respect to the total number of resources used at the input and 
the detection process?
As our last result, we prove that the answer is affirmative 
by showing a concrete measurement scheme.

%%%%%%%%%%%%%%%%%%%%%%%%%%%%%%%%%%%%%%%%%%%%
\begin{figure}
\begin{center}
\includegraphics[width=90mm]{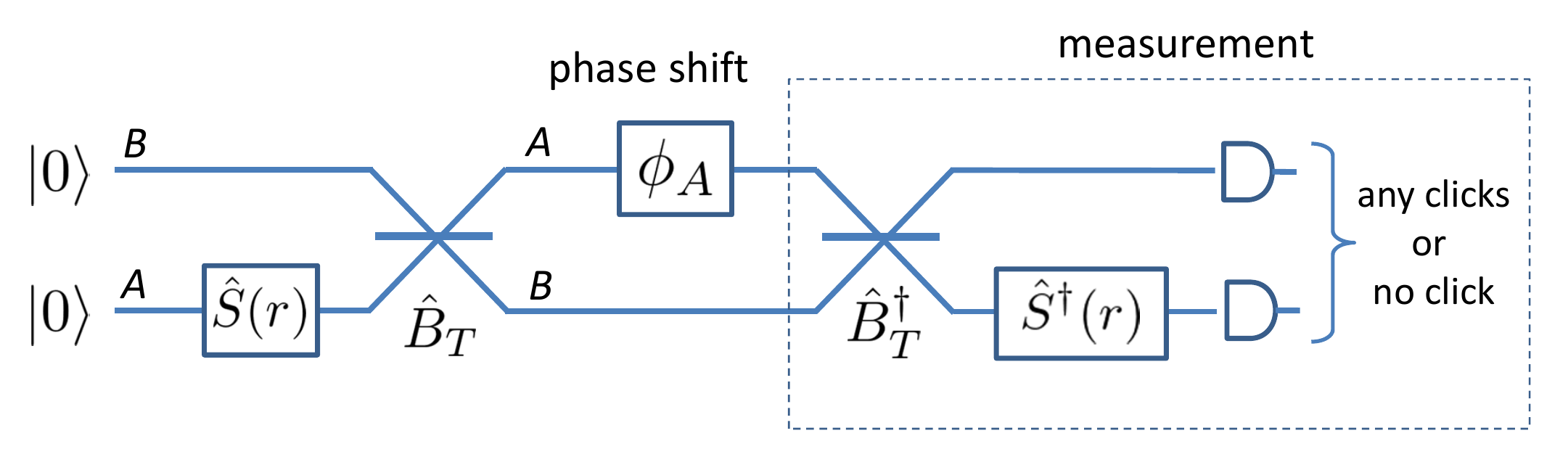}   %
\caption{\label{fig:phase-sensitive_meas}
Mach-Zehnder interferometer sensing with 
the input-phase sensitive measurement. 
$\hat{S}(r)$ is the squeezing operator with squeezing parameter $r$ 
and $\hat{B}_T$ is the beam-splitting operator with transmissivity $T$. 
}
\end{center}
\end{figure}
%%%%%%%%%%%%%%%%%%%%%%%%%%%%%%%%%%%%%%%%%%%%

Figure \ref{fig:phase-sensitive_meas} illustrates the concrete input state 
and the measurement. 
The input state is a single-mode squeezed vacuum, which is generated 
from vacuum by applying the squeezing operation $\hat{S}(r)$, where 
$r$ is the squeezing parameter.
The measurement is a time-reversed process, that is, it 
consists of the complex-conjugate beam splitter $\hat{B}_T^\dagger$, 
anti-squeezing operation $\hat{S}^\dagger(r)$, and photon detectors that 
discriminate zero and non-zero photons (so-called on-off detectors). 
For simplicity of the analysis, we consider only two outcomes: 
zero photons in both detectors or other events, 
\{$|0\rangle\langle0| \otimes |0\rangle\langle0|, 
I - |0\rangle\langle0| \otimes |0\rangle\langle0|$\} 
(photon number discrimination may further improve the performance but 
to simplify the discussion here, we leave it for a future work). 
Note that this mirror-image like detection strategy has been considered in 
the context of state discrimination \cite{TBS03,TS08}
and also the phase estimation via coherent-state input \cite{ITWFES16}. 
Since we need the phase information of the input state at 
the anti-squeezing process, this is a {\it phase-sensitive} measurement, 
and so the POVM has access to external phase information and the phase of the squeezer $\hat{S}^\dagger$. 
The input average photon number is given by $\bar{n}= \sinh^2 r$. 
Since the measurement device also uses the same amount of squeezing, 
the average photon number of the all resources are counted as 
$\bar{n}_{\rm tot} = 2\bar{n}$ (to consider the fundamental limitation, 
here we assume a unit efficiency parametric downconverter where 
$\sinh^2 r$ is directly a function of the pump energy used for the converter).

The attainable precision limit (in the asymptotic limit, $r\rightarrow\infty$) is specified by 
calculating its CFI 
\cite{HelstromBook,HayashiBook}: 
%%%%%%%%%%%%%%%%%%%%%%%%%%%
\begin{equation}
\label{eq:CFI}
F(\phi_A) 
= {\rm E}\left[ - \frac{d^2}{d \phi_A^2} \log p_{\phi_A} \right], 
\end{equation}
%%%%%%%%%%%%%%%%%%%%%%%%%%%
where $p_{\phi_A} = P(x|\phi_A)$ is the conditional probability of obtaining 
the measurement outcome $x$ for given $\phi_A$ and 
$x=0,1$ represent the photon detection outcome 
$|0\rangle\langle0|\otimes|0\rangle\langle0|$ and 
$I-|0\rangle\langle0|\otimes|0\rangle\langle0|$, respectively.

$F(\phi_A)$ is calculated by the characteristic function approach 
(e.g. Ref.~\cite{TJS15}), and the derived analytical expression 
of $F(\phi_A)$ is complicated (see Supplementary Material 4).
Taking the limit of $\phi_A \to 0$, we get 
%%%%%%%%%%%%%%%%%%%%%%%%%%%
\begin{equation}
\label{eq:CFI_phi_A}
F(\phi_A) = 2\bar{n}_{\rm tot} T (1+T+\bar{n}_{\rm tot}T), 
\end{equation}
%%%%%%%%%%%%%%%%%%%%%%%%%%%
where we remind the reader that $\bar{n}_{\rm tot} = 2\bar{n}$ is the \textit{total} resource 
used for the input state and the detection process. 
Thus we get the (classical) CRB around $\phi_A = 0$ as 
%%%%%%%%%%%%%%%%%%%%%%%%%%%
\begin{equation}
\label{eq:CCRB}
\Delta^2\phi_A \ge 
\frac{1}{2m\bar{n}_{\rm tot} T (1+T+\bar{n}_{\rm tot}T)}, 
\end{equation}
%%%%%%%%%%%%%%%%%%%%%%%%%%%
which surpasses the SNL of the total resource for any $T \ne 0$. 
This example shows how a QFI-only calculation could contain hidden 
resources in the unknown optimal POVM, that are unfairly not counted.
Here, by taking all resources into account, we conclude that it is possible to beat the SNL 
for the $\hat{g}_1$ estimation if one uses an additional energy and phase resources 
at the detection.

{\it Conclusions.---} 
In this paper, we revisit the ultimate limit of the MZI sensing precision 
when an input into one port is vacuum. 
We show a full statement of the problem with a rigorous proof: 
the statement depends on your choice of the phase shift unitary operator---
in other words, on the physical setup of your sensing application. 

First, if both two arms of the MZI have different unknown phase shifts 
in your application (e.g. gravitational wave detection) and 
input vacuum in one port, then no matter what you put in the other port, 
and no matter what your detection scheme you deploy, you can never 
do better than the SNL in phase sensitivity.
The statement holds even if the first beamsplitter of the MZI is 
non-50:50. 
The proof is based on the fact that it is intrinsically a two-parameter estimation problem 
although the phase sum $\phi_s$ is often treated as a ``global phase'' and ignored in real experiments. 
Intuitively, we see that the state after the phase shift is $e^{i \hat{g}_s \phi_s} e^{i \hat{g}_d \phi_d} |\Psi\rangle_{AB}$ 
(where $|\Psi\rangle_{AB}$ is a probe state before the phase shift) which clearly contains $\phi_s$. 
This implies if one knows nothing about $\phi_s$, the effective state should be randomized over $\phi_s$ 
which might give a different conclusion than that from the analysis considering only $\phi_d$ as a single-parameter estimation. 
This is why we need to take into account both $\phi_s$ and $\phi_d$ even for deriving the precision bound of only $\phi_d$. 
This type of sensing includes the gravitational wave detection 
\cite{GEO600,GEO6001}, long-baseline interferometry \cite{PKG16}, 
and differential interference contrast microscopy \cite{N55,OOT13}, for example. 
In these applications, if one input is vacuum, our result rules out 
the possibility of doing something ``quantum'' at the detector 
(such as putting in a squeezer or doing photon addition or subtraction) 
to beat the SNL.

Second, if only one of the MZI arms has an unknown phase shift 
in your application (sensing a sample placed at one arm of the MZI), 
the ultimate precision limit depends on the detector restriction. 
If you do not allow the detector to use any external phase reference and 
power resource, then the precision is limited by the SNL. 
However, if you allow the detector to use such resources, 
you can beat the SNL in terms of 
the total resource used at the input and detector. 
The explicit sensing scheme which uses squeezers for both input and detector 
is given. 
This type of sensing includes simple MZI devices measuring sample's density, 
pressure, temperature, etc, and also LIDAR-type sensing \cite{GLY16}. 
In these applications, only if you put nonclassical light into at least one input port, 
is there a hope to beat the SNL by doing something quantum 
at the detector, even if the other port is vacuum.

\section{Acknowledgements} 
MT would like to acknowledge suport from the Open Partnership Joint Projects of JSPS Bilateral Joint
Research Projects and the ImPACT Program of Council
for Science, Technology and Innovation, Japan.
KPS, and JPD would like to acknowledge support from the 
Air Force Office of Scientific Research, the Army Research Office, the National Science 
Foundation, and the Northrop Grumman Corporation. CY would like to acknowledge support 
from an Economic Development Assistantship from the the Louisiana State University System Board of Regents.

%\section*{References}

\bibliographystyle{apsrev4-1}
\bibliography{ref1}

\begin{thebibliography}{25}%
\makeatletter
\providecommand \@ifxundefined [1]{%
 \@ifx{#1\undefined}
}%
\providecommand \@ifnum [1]{%
 \ifnum #1\expandafter \@firstoftwo
 \else \expandafter \@secondoftwo
 \fi
}%
\providecommand \@ifx [1]{%
 \ifx #1\expandafter \@firstoftwo
 \else \expandafter \@secondoftwo
 \fi
}%
\providecommand \natexlab [1]{#1}%
\providecommand \enquote  [1]{``#1''}%
\providecommand \bibnamefont  [1]{#1}%
\providecommand \bibfnamefont [1]{#1}%
\providecommand \citenamefont [1]{#1}%
\providecommand \href@noop [0]{\@secondoftwo}%
\providecommand \href [0]{\begingroup \@sanitize@url \@href}%
\providecommand \@href[1]{\@@startlink{#1}\@@href}%
\providecommand \@@href[1]{\endgroup#1\@@endlink}%
\providecommand \@sanitize@url [0]{\catcode `\\12\catcode `\$12\catcode
  `\&12\catcode `\#12\catcode `\^12\catcode `\_12\catcode `\%12\relax}%
\providecommand \@@startlink[1]{}%
\providecommand \@@endlink[0]{}%
\providecommand \url  [0]{\begingroup\@sanitize@url \@url }%
\providecommand \@url [1]{\endgroup\@href {#1}{\urlprefix }}%
\providecommand \urlprefix  [0]{URL }%
\providecommand \Eprint [0]{\href }%
\providecommand \doibase [0]{http://dx.doi.org/}%
\providecommand \selectlanguage [0]{\@gobble}%
\providecommand \bibinfo  [0]{\@secondoftwo}%
\providecommand \bibfield  [0]{\@secondoftwo}%
\providecommand \translation [1]{[#1]}%
\providecommand \BibitemOpen [0]{}%
\providecommand \bibitemStop [0]{}%
\providecommand \bibitemNoStop [0]{.\EOS\space}%
\providecommand \EOS [0]{\spacefactor3000\relax}%
\providecommand \BibitemShut  [1]{\csname bibitem#1\endcsname}%
\let\auto@bib@innerbib\@empty
%</preamble>
\bibitem [{\citenamefont {Giovannetti}\ \emph {et~al.}(2011)\citenamefont
  {Giovannetti}, \citenamefont {Lloyd},\ and\ \citenamefont {Maccone}}]{GLM11}%
  \BibitemOpen
  \bibfield  {author} {\bibinfo {author} {\bibfnamefont {V.}~\bibnamefont
  {Giovannetti}}, \bibinfo {author} {\bibfnamefont {S.}~\bibnamefont {Lloyd}},
  \ and\ \bibinfo {author} {\bibfnamefont {L.}~\bibnamefont {Maccone}},\ }\href
  {\doibase 10.1038/nphoton.2011.35} {\bibfield  {journal} {\bibinfo  {journal}
  {Nat. Photon.}\ }\textbf {\bibinfo {volume} {5}},\ \bibinfo {pages} {222}
  (\bibinfo {year} {2011})}\BibitemShut {NoStop}%
\bibitem [{\citenamefont {Demkowicz-Dobrza{\`n}ski}\ \emph
  {et~al.}(2015)\citenamefont {Demkowicz-Dobrza{\`n}ski}, \citenamefont
  {Jarzyna},\ and\ \citenamefont {Ko{\l}ody{\`n}ski}}]{DJK15}%
  \BibitemOpen
  \bibfield  {author} {\bibinfo {author} {\bibfnamefont {R.}~\bibnamefont
  {Demkowicz-Dobrza{\`n}ski}}, \bibinfo {author} {\bibfnamefont
  {M.}~\bibnamefont {Jarzyna}}, \ and\ \bibinfo {author} {\bibfnamefont
  {J.}~\bibnamefont {Ko{\l}ody{\`n}ski}},\ }\href {\doibase
  http://dx.doi.org/10.1016/bs.po.2015.02.003} {\emph {\bibinfo {title}
  {Quantum Limits in Optical Interferometry}}},\ edited by\ \bibinfo {editor}
  {\bibfnamefont {E.}~\bibnamefont {Wolf}},\ \bibinfo {series} {Progress in
  Optics}, Vol.~\bibinfo {volume} {60}\ (\bibinfo  {publisher} {Elsevier},\
  \bibinfo {year} {2015})\ pp.\ \bibinfo {pages} {345 -- 435}\BibitemShut
  {NoStop}%
\bibitem [{\citenamefont {Dowling}\ and\ \citenamefont
  {Seshadreesan}(2015)}]{DS15}%
  \BibitemOpen
  \bibfield  {author} {\bibinfo {author} {\bibfnamefont {J.~P.}\ \bibnamefont
  {Dowling}}\ and\ \bibinfo {author} {\bibfnamefont {K.~P.}\ \bibnamefont
  {Seshadreesan}},\ }\href {\doibase 10.1109/JLT.2014.2386795} {\bibfield
  {journal} {\bibinfo  {journal} {J. Lightwave Technol.}\ }\textbf {\bibinfo
  {volume} {33}},\ \bibinfo {pages} {2359} (\bibinfo {year}
  {2015})}\BibitemShut {NoStop}%
\bibitem [{\citenamefont {Caves}(1981)}]{caves81}%
  \BibitemOpen
  \bibfield  {author} {\bibinfo {author} {\bibfnamefont {C.~M.}\ \bibnamefont
  {Caves}},\ }\href {\doibase 10.1103/PhysRevD.23.1693} {\bibfield  {journal}
  {\bibinfo  {journal} {Phys. Rev. D}\ }\textbf {\bibinfo {volume} {23}},\
  \bibinfo {pages} {1693} (\bibinfo {year} {1981})}\BibitemShut {NoStop}%
\bibitem [{\citenamefont {Grote}\ \emph {et~al.}(2013)\citenamefont {Grote},
  \citenamefont {Danzmann}, \citenamefont {Dooley}, \citenamefont {Schnabel},
  \citenamefont {Slutsky},\ and\ \citenamefont {Vahlbruch}}]{GEO600}%
  \BibitemOpen
  \bibfield  {author} {\bibinfo {author} {\bibfnamefont {H.}~\bibnamefont
  {Grote}}, \bibinfo {author} {\bibfnamefont {K.}~\bibnamefont {Danzmann}},
  \bibinfo {author} {\bibfnamefont {K.~L.}\ \bibnamefont {Dooley}}, \bibinfo
  {author} {\bibfnamefont {R.}~\bibnamefont {Schnabel}}, \bibinfo {author}
  {\bibfnamefont {J.}~\bibnamefont {Slutsky}}, \ and\ \bibinfo {author}
  {\bibfnamefont {H.}~\bibnamefont {Vahlbruch}},\ }\href {\doibase
  10.1103/PhysRevLett.110.181101} {\bibfield  {journal} {\bibinfo  {journal}
  {Phys. Rev. Lett.}\ }\textbf {\bibinfo {volume} {110}},\ \bibinfo {pages}
  {181101} (\bibinfo {year} {2013})}\BibitemShut {NoStop}%
\bibitem [{\citenamefont {Lugiato}\ \emph {et~al.}(2002)\citenamefont
  {Lugiato}, \citenamefont {Gatti},\ and\ \citenamefont {Brambilla}}]{GEO6001}%
  \BibitemOpen
  \bibfield  {author} {\bibinfo {author} {\bibfnamefont {L.~A.}\ \bibnamefont
  {Lugiato}}, \bibinfo {author} {\bibfnamefont {A.}~\bibnamefont {Gatti}}, \
  and\ \bibinfo {author} {\bibfnamefont {E.}~\bibnamefont {Brambilla}},\ }\href
  {http://stacks.iop.org/1464-4266/4/i=3/a=372} {\bibfield  {journal} {\bibinfo
   {journal} {J. Opt. B: Quantum Semiclass. Opt.}\ }\textbf {\bibinfo {volume}
  {4}},\ \bibinfo {pages} {S176} (\bibinfo {year} {2002})}\BibitemShut
  {NoStop}%
\bibitem [{\citenamefont {Lang}\ and\ \citenamefont {Caves}(2013)}]{LC13}%
  \BibitemOpen
  \bibfield  {author} {\bibinfo {author} {\bibfnamefont {M.~D.}\ \bibnamefont
  {Lang}}\ and\ \bibinfo {author} {\bibfnamefont {C.~M.}\ \bibnamefont
  {Caves}},\ }\href {\doibase 10.1103/PhysRevLett.111.173601} {\bibfield
  {journal} {\bibinfo  {journal} {Phys. Rev. Lett.}\ }\textbf {\bibinfo
  {volume} {111}},\ \bibinfo {pages} {173601} (\bibinfo {year}
  {2013})}\BibitemShut {NoStop}%
\bibitem [{\citenamefont {Jarzyna}\ and\ \citenamefont
  {Demkowicz-Dobrza\ifmmode~\acute{n}\else \'{n}\fi{}ski}(2012)}]{JD12}%
  \BibitemOpen
  \bibfield  {author} {\bibinfo {author} {\bibfnamefont {M.}~\bibnamefont
  {Jarzyna}}\ and\ \bibinfo {author} {\bibfnamefont {R.}~\bibnamefont
  {Demkowicz-Dobrza\ifmmode~\acute{n}\else \'{n}\fi{}ski}},\ }\href {\doibase
  10.1103/PhysRevA.85.011801} {\bibfield  {journal} {\bibinfo  {journal} {Phys.
  Rev. A}\ }\textbf {\bibinfo {volume} {85}},\ \bibinfo {pages} {011801}
  (\bibinfo {year} {2012})}\BibitemShut {NoStop}%
\bibitem [{\citenamefont {Pezz\`e}\ \emph {et~al.}(2015)\citenamefont
  {Pezz\`e}, \citenamefont {Hyllus},\ and\ \citenamefont {Smerzi}}]{PHS15}%
  \BibitemOpen
  \bibfield  {author} {\bibinfo {author} {\bibfnamefont {L.}~\bibnamefont
  {Pezz\`e}}, \bibinfo {author} {\bibfnamefont {P.}~\bibnamefont {Hyllus}}, \
  and\ \bibinfo {author} {\bibfnamefont {A.}~\bibnamefont {Smerzi}},\ }\href
  {\doibase 10.1103/PhysRevA.91.032103} {\bibfield  {journal} {\bibinfo
  {journal} {Phys. Rev. A}\ }\textbf {\bibinfo {volume} {91}},\ \bibinfo
  {pages} {032103} (\bibinfo {year} {2015})}\BibitemShut {NoStop}%
\bibitem [{\citenamefont {Helstrom}(1976)}]{HelstromBook}%
  \BibitemOpen
  \bibfield  {author} {\bibinfo {author} {\bibfnamefont {C.~W.}\ \bibnamefont
  {Helstrom}},\ }\href@noop {} {\emph {\bibinfo {title} {Quantum Detection and
  Estimation Theory}}}\ (\bibinfo  {publisher} {Academic Press, New York},\
  \bibinfo {year} {1976})\BibitemShut {NoStop}%
\bibitem [{\citenamefont {Anisimov}\ \emph {et~al.}(2010)\citenamefont
  {Anisimov}, \citenamefont {Raterman}, \citenamefont {Chiruvelli},
  \citenamefont {Plick}, \citenamefont {Huver}, \citenamefont {Lee},\ and\
  \citenamefont {Dowling}}]{ARCPHLD10}%
  \BibitemOpen
  \bibfield  {author} {\bibinfo {author} {\bibfnamefont {P.~M.}\ \bibnamefont
  {Anisimov}}, \bibinfo {author} {\bibfnamefont {G.~M.}\ \bibnamefont
  {Raterman}}, \bibinfo {author} {\bibfnamefont {A.}~\bibnamefont
  {Chiruvelli}}, \bibinfo {author} {\bibfnamefont {W.~N.}\ \bibnamefont
  {Plick}}, \bibinfo {author} {\bibfnamefont {S.~D.}\ \bibnamefont {Huver}},
  \bibinfo {author} {\bibfnamefont {H.}~\bibnamefont {Lee}}, \ and\ \bibinfo
  {author} {\bibfnamefont {J.~P.}\ \bibnamefont {Dowling}},\ }\href {\doibase
  10.1103/PhysRevLett.104.103602} {\bibfield  {journal} {\bibinfo  {journal}
  {Phys. Rev. Lett.}\ }\textbf {\bibinfo {volume} {104}},\ \bibinfo {pages}
  {103602} (\bibinfo {year} {2010})}\BibitemShut {NoStop}%
\bibitem [{\citenamefont {Nomarski}(1955)}]{N55}%
  \BibitemOpen
  \bibfield  {author} {\bibinfo {author} {\bibfnamefont {M.}~\bibnamefont
  {Nomarski}},\ }\href@noop {} {\bibfield  {journal} {\bibinfo  {journal}
  {Journal de Physique et le Radium}\ }\textbf {\bibinfo {volume} {16}},\
  \bibinfo {pages} {S9} (\bibinfo {year} {1955})}\BibitemShut {NoStop}%
\bibitem [{\citenamefont {Ono}\ \emph {et~al.}(2013)\citenamefont {Ono},
  \citenamefont {Okamoto},\ and\ \citenamefont {Takeuchi}}]{OOT13}%
  \BibitemOpen
  \bibfield  {author} {\bibinfo {author} {\bibfnamefont {T.}~\bibnamefont
  {Ono}}, \bibinfo {author} {\bibfnamefont {R.}~\bibnamefont {Okamoto}}, \ and\
  \bibinfo {author} {\bibfnamefont {S.}~\bibnamefont {Takeuchi}},\ }\href
  {\doibase 10.1038/ncomms3426} {\bibfield  {journal} {\bibinfo  {journal}
  {Nat. Commun.}\ }\textbf {\bibinfo {volume} {4}},\ \bibinfo {pages} {2426}
  (\bibinfo {year} {2013})}\BibitemShut {NoStop}%
\bibitem [{Note1()}]{Note1}%
  \BibitemOpen
  \bibinfo {note} {In Ref.~\cite {LC13}, the QFIM of the system considered was
  calculated. However, they reduce it to the single-parameter estimation (i.e.
  drop off the terms for $\phi _s$) which looses the tightness of the bound.
  Note that this problem does not appear in Eq.~(\ref {eq:qfi_LC13}) since with
  $T=1/2$, the non-diagonal term of the QFIM goes to zero and thus the problem
  reduces to two independent single-parameter estimations. Nevertheless, in
  Ref.~\cite {LC13}, they also consider the non-vacuum input case where the
  bound may have some looseness.}\BibitemShut {Stop}%
\bibitem [{\citenamefont {Petz}(1996)}]{P96}%
  \BibitemOpen
  \bibfield  {author} {\bibinfo {author} {\bibfnamefont {D.}~\bibnamefont
  {Petz}},\ }\href {\doibase 10.1016/0024-3795(94)00211-8} {\bibfield
  {journal} {\bibinfo  {journal} {Linear Algebra and Its Applications}\
  }\textbf {\bibinfo {volume} {244}},\ \bibinfo {pages} {81} (\bibinfo {year}
  {1996})}\BibitemShut {NoStop}%
\bibitem [{\citenamefont {Petz}\ and\ \citenamefont {Ghinea}(2010)}]{PG10}%
  \BibitemOpen
  \bibfield  {author} {\bibinfo {author} {\bibfnamefont {D.}~\bibnamefont
  {Petz}}\ and\ \bibinfo {author} {\bibfnamefont {C.}~\bibnamefont {Ghinea}},\
  }\href {http://arxiv.org/abs/1008.2417} {\ \textbf {\bibinfo {volume} {27}},\
  \bibinfo {pages} {261} (\bibinfo {year} {2010})}\BibitemShut {NoStop}%
\bibitem [{\citenamefont {Fujiwara}(2001)}]{F01}%
  \BibitemOpen
  \bibfield  {author} {\bibinfo {author} {\bibfnamefont {A.}~\bibnamefont
  {Fujiwara}},\ }\href {\doibase 10.1103/PhysRevA.63.042304} {\bibfield
  {journal} {\bibinfo  {journal} {Phys. Rev. A}\ }\textbf {\bibinfo {volume}
  {63}},\ \bibinfo {pages} {042304} (\bibinfo {year} {2001})}\BibitemShut
  {NoStop}%
\bibitem [{\citenamefont {Takeoka}\ \emph {et~al.}(2003)\citenamefont
  {Takeoka}, \citenamefont {Ban},\ and\ \citenamefont {Sasaki}}]{TBS03}%
  \BibitemOpen
  \bibfield  {author} {\bibinfo {author} {\bibfnamefont {M.}~\bibnamefont
  {Takeoka}}, \bibinfo {author} {\bibfnamefont {M.}~\bibnamefont {Ban}}, \ and\
  \bibinfo {author} {\bibfnamefont {M.}~\bibnamefont {Sasaki}},\ }\href
  {\doibase 10.1103/PhysRevA.68.012307} {\bibfield  {journal} {\bibinfo
  {journal} {Phys. Rev. A}\ }\textbf {\bibinfo {volume} {68}},\ \bibinfo
  {pages} {012307} (\bibinfo {year} {2003})}\BibitemShut {NoStop}%
\bibitem [{\citenamefont {Takeoka}\ and\ \citenamefont {Sasaki}(2008)}]{TS08}%
  \BibitemOpen
  \bibfield  {author} {\bibinfo {author} {\bibfnamefont {M.}~\bibnamefont
  {Takeoka}}\ and\ \bibinfo {author} {\bibfnamefont {M.}~\bibnamefont
  {Sasaki}},\ }\href {\doibase 10.1103/PhysRevA.78.022320} {\bibfield
  {journal} {\bibinfo  {journal} {Phys. Rev. A}\ }\textbf {\bibinfo {volume}
  {78}},\ \bibinfo {pages} {022320} (\bibinfo {year} {2008})}\BibitemShut
  {NoStop}%
\bibitem [{\citenamefont {Izumi}\ \emph {et~al.}(2016)\citenamefont {Izumi},
  \citenamefont {Takeoka}, \citenamefont {Wakui}, \citenamefont {Fujiwara},
  \citenamefont {Ema},\ and\ \citenamefont {Sasaki}}]{ITWFES16}%
  \BibitemOpen
  \bibfield  {author} {\bibinfo {author} {\bibfnamefont {S.}~\bibnamefont
  {Izumi}}, \bibinfo {author} {\bibfnamefont {M.}~\bibnamefont {Takeoka}},
  \bibinfo {author} {\bibfnamefont {K.}~\bibnamefont {Wakui}}, \bibinfo
  {author} {\bibfnamefont {M.}~\bibnamefont {Fujiwara}}, \bibinfo {author}
  {\bibfnamefont {K.}~\bibnamefont {Ema}}, \ and\ \bibinfo {author}
  {\bibfnamefont {M.}~\bibnamefont {Sasaki}},\ }\href {\doibase
  10.1103/PhysRevA.94.033842} {\bibfield  {journal} {\bibinfo  {journal} {Phys.
  Rev. A}\ }\textbf {\bibinfo {volume} {94}},\ \bibinfo {pages} {033842}
  (\bibinfo {year} {2016})}\BibitemShut {NoStop}%
\bibitem [{\citenamefont {Hayashi}(2006)}]{HayashiBook}%
  \BibitemOpen
  \bibfield  {author} {\bibinfo {author} {\bibfnamefont {M.}~\bibnamefont
  {Hayashi}},\ }\href@noop {} {\emph {\bibinfo {title} {Quantum Information: An
  Introduction}}}\ (\bibinfo  {publisher} {Springer},\ \bibinfo {year}
  {2006})\BibitemShut {NoStop}%
\bibitem [{\citenamefont {Takeoka}\ \emph {et~al.}(2015)\citenamefont
  {Takeoka}, \citenamefont {Jin},\ and\ \citenamefont {Sasaki}}]{TJS15}%
  \BibitemOpen
  \bibfield  {author} {\bibinfo {author} {\bibfnamefont {M.}~\bibnamefont
  {Takeoka}}, \bibinfo {author} {\bibfnamefont {R.}~\bibnamefont {Jin}}, \ and\
  \bibinfo {author} {\bibfnamefont {M.}~\bibnamefont {Sasaki}},\ }\href
  {\doibase 10.1088/1367-2630/17/4/043030} {\bibfield  {journal} {\bibinfo
  {journal} {New J. Phys.}\ }\textbf {\bibinfo {volume} {17}},\ \bibinfo
  {pages} {43030} (\bibinfo {year} {2015})}\BibitemShut {NoStop}%
\bibitem [{\citenamefont {Parazzoli}\ \emph {et~al.}(2016)\citenamefont
  {Parazzoli}, \citenamefont {Koltenbah}, \citenamefont {Gerwe}, \citenamefont
  {Idell}, \citenamefont {Gard}, \citenamefont {Birrittella}, \citenamefont
  {Rafsanjani}, \citenamefont {Mirhosseini}, \citenamefont {Magan-Loiza},
  \citenamefont {Dowling} \emph {et~al.}}]{PKG16}%
  \BibitemOpen
  \bibfield  {author} {\bibinfo {author} {\bibfnamefont {C.}~\bibnamefont
  {Parazzoli}}, \bibinfo {author} {\bibfnamefont {B.}~\bibnamefont
  {Koltenbah}}, \bibinfo {author} {\bibfnamefont {D.}~\bibnamefont {Gerwe}},
  \bibinfo {author} {\bibfnamefont {P.}~\bibnamefont {Idell}}, \bibinfo
  {author} {\bibfnamefont {B.}~\bibnamefont {Gard}}, \bibinfo {author}
  {\bibfnamefont {R.}~\bibnamefont {Birrittella}}, \bibinfo {author}
  {\bibfnamefont {S.~M.}\ \bibnamefont {Rafsanjani}}, \bibinfo {author}
  {\bibfnamefont {M.}~\bibnamefont {Mirhosseini}}, \bibinfo {author}
  {\bibfnamefont {O.~S.}\ \bibnamefont {Magan-Loiza}}, \bibinfo {author}
  {\bibfnamefont {J.}~\bibnamefont {Dowling}},  \emph {et~al.},\ }\href
  {http://arxiv.org/abs/1609.02780} {\bibfield  {journal} {\bibinfo  {journal}
  {arXiv preprint arXiv:1609.02780}\ } (\bibinfo {year} {2016})}\BibitemShut
  {NoStop}%
\bibitem [{\citenamefont {Gard}\ \emph {et~al.}(2016)\citenamefont {Gard},
  \citenamefont {Li}, \citenamefont {You}, \citenamefont {Seshadreesan},
  \citenamefont {Birrittella}, \citenamefont {Luine}, \citenamefont
  {Rafsanjani}, \citenamefont {Mirhosseini}, \citenamefont {Maga{\~n}a-Loaiza},
  \citenamefont {Koltenbah} \emph {et~al.}}]{GLY16}%
  \BibitemOpen
  \bibfield  {author} {\bibinfo {author} {\bibfnamefont {B.~T.}\ \bibnamefont
  {Gard}}, \bibinfo {author} {\bibfnamefont {D.}~\bibnamefont {Li}}, \bibinfo
  {author} {\bibfnamefont {C.}~\bibnamefont {You}}, \bibinfo {author}
  {\bibfnamefont {K.~P.}\ \bibnamefont {Seshadreesan}}, \bibinfo {author}
  {\bibfnamefont {R.}~\bibnamefont {Birrittella}}, \bibinfo {author}
  {\bibfnamefont {J.}~\bibnamefont {Luine}}, \bibinfo {author} {\bibfnamefont
  {S.~M.~H.}\ \bibnamefont {Rafsanjani}}, \bibinfo {author} {\bibfnamefont
  {M.}~\bibnamefont {Mirhosseini}}, \bibinfo {author} {\bibfnamefont {O.~S.}\
  \bibnamefont {Maga{\~n}a-Loaiza}}, \bibinfo {author} {\bibfnamefont {B.~E.}\
  \bibnamefont {Koltenbah}},  \emph {et~al.},\ }\href
  {http://arxiv.org/abs/1609.09598} {\bibfield  {journal} {\bibinfo  {journal}
  {arXiv preprint arXiv:1606.09598}\ } (\bibinfo {year} {2016})}\BibitemShut
  {NoStop}%
\bibitem [{\citenamefont {Weedbrook}\ \emph {et~al.}(2012)\citenamefont
  {Weedbrook}, \citenamefont {Pirandola}, \citenamefont
  {Garc{\'{i}}a-Patr{\'{o}}n}, \citenamefont {Cerf}, \citenamefont {Ralph},
  \citenamefont {Shapiro},\ and\ \citenamefont {Lloyd}}]{WPGCRSL12}%
  \BibitemOpen
  \bibfield  {author} {\bibinfo {author} {\bibfnamefont {C.}~\bibnamefont
  {Weedbrook}}, \bibinfo {author} {\bibfnamefont {S.}~\bibnamefont
  {Pirandola}}, \bibinfo {author} {\bibfnamefont {R.}~\bibnamefont
  {Garc{\'{i}}a-Patr{\'{o}}n}}, \bibinfo {author} {\bibfnamefont
  {N.}~\bibnamefont {Cerf}}, \bibinfo {author} {\bibfnamefont {T.}~\bibnamefont
  {Ralph}}, \bibinfo {author} {\bibfnamefont {J.}~\bibnamefont {Shapiro}}, \
  and\ \bibinfo {author} {\bibfnamefont {S.}~\bibnamefont {Lloyd}},\ }\href
  {\doibase 10.1103/RevModPhys.84.621} {\bibfield  {journal} {\bibinfo
  {journal} {Rev. Mod. Phys.}\ }\textbf {\bibinfo {volume} {84}},\ \bibinfo
  {pages} {621} (\bibinfo {year} {2012})}\BibitemShut {NoStop}%
\end{thebibliography}%

\appendix

\section{Supplemental Material 1: Quantum Fisher information 
for the Mach-Zehnder interferometer phase sensing with a vacuum input}

Here we derive Eqs.~(\ref{eq:qfi_g_1}), (\ref{eq:qfi_g_d}), and 
(\ref{eq:QFIM_entry_dd})--(\ref{eq:QFIM_entry_ds}) in the main text. 
Consider $|\chi\rangle \otimes |0\rangle$ as an input 
to the MZ interferometer. For the calculation, it is useful to expand 
$|\chi\rangle$ in a coherent state basis: 
%%%%%%%%%%%%%%%%%%%%%%%%%%%
\begin{equation}
\label{eq:pure_input}
|\chi\rangle = \int {\rm d}^2 \alpha \, f(\alpha) |\alpha\rangle ,
\end{equation}
%%%%%%%%%%%%%%%%%%%%%%%%%%%
where $|\alpha\rangle$ is a coherent state with complex quadrature 
amplitude $\alpha$. 
Then the average photon number and the variance of the state are 
given by 
%%%%%%%%%%%%%%%%%%%%%%%%%%%
\begin{eqnarray}
\label{eq:<n>}
\bar{n}_\chi & = & \langle\chi|\hat{n}|\chi\rangle 
\nonumber\\ & = & 
\int {\rm d}^2 \alpha \, \int {\rm d}^2 \beta \, f^*(\alpha) f(\beta) 
\langle\alpha| \hat{n} |\beta\rangle 
\nonumber\\ & = & 
\int {\rm d}^2 \alpha \, \int {\rm d}^2 \beta \, f^*(\alpha) f(\beta) 
\alpha^* \beta \langle\alpha|\beta\rangle
\nonumber\\ & = & 
\int {\rm d}^2 \alpha \, \int {\rm d}^2 \beta \, f^*(\alpha) f(\beta) 
\alpha^* \beta 
\nonumber\\ && \times 
\exp\left[ -\frac{1}{2} \left( |\alpha|^2 + |\beta|^2 - 2 \alpha^* \beta 
\right) \right] , 
\end{eqnarray}
%%%%%%%%%%%%%%%%%%%%%%%%%%%
and
%%%%%%%%%%%%%%%%%%%%%%%%%%%
\begin{eqnarray}
\label{eq:<n^2>}
V_\chi & = & \langle\chi|\hat{n}^2|\chi\rangle - \bar{n}_\chi^2
\nonumber\\ & = & 
\int {\rm d}^2 \alpha \, \int {\rm d}^2 \beta \, f^*(\alpha) f(\beta) 
\left\{ (\alpha^* \beta)^2 + \alpha^* \beta \right\} 
\nonumber\\ && \times 
\exp\left[ -\frac{1}{2} \left( |\alpha|^2 + |\beta|^2 - 2 \alpha^* \beta 
\right) \right]  - \bar{n}_\chi^2, 
\end{eqnarray}
%%%%%%%%%%%%%%%%%%%%%%%%%%%
where we use the fact that 
$\hat{n}^2 = \hat{a}^{\dagger \, 2} \hat{a}^2 + \hat{a}^\dagger \hat{a}$.

The state after the beam splitter with transmittance $T$ 
is given by 
%%%%%%%%%%%%%%%%%%%%%%%%%%%
\begin{equation}
\label{eq:Phi}
|\Phi\rangle_{AB} = 
\int {\rm d}^2 \alpha \, f(\alpha) 
\left|\sqrt{T}\alpha\right\rangle_A 
\left|\sqrt{R}\alpha\right\rangle_B ,
\end{equation}
%%%%%%%%%%%%%%%%%%%%%%%%%%%
where $R=1-T$.

\subsection{QFI with $\hat{g}_1$ (Eq.~(\ref{eq:qfi_g_1}))}
The quantum Fisher information (QFI) is calculated from  
%%%%%%%%%%%%%%%%%%%%%%%%%%%
\begin{equation}
F_Q(|\chi\rangle, \hat{g}_1, T) = 4 (\langle\Phi| \hat{g}_1^2 |\Phi\rangle 
- \langle\Phi| \hat{g}_1 |\Phi\rangle^2) .
\end{equation}
%%%%%%%%%%%%%%%%%%%%%%%%%%%
We have 
\begin{widetext}
%%%%%%%%%%%%%%%%%%%%%%%%%%%
\begin{eqnarray}
\langle\Phi| \hat{g}_1^2 |\Phi\rangle & = & 
\int {\rm d}^2 \alpha \, \int {\rm d}^2 \beta \, f^*(\alpha) f(\beta) 
\left\langle \sqrt{T}\alpha \right|_A  
\left\langle \sqrt{R}\alpha \right|_B  
\left( 
\hat{a}^{\dagger \, 2} \hat{a}^2 + \hat{a}^\dagger \hat{a} 
\right)
\left| \sqrt{T}\beta \right\rangle_A
\left| \sqrt{R}\beta \right\rangle_B
\nonumber\\ & = & 
\int {\rm d}^2 \alpha \, \int {\rm d}^2 \beta \, f^*(\alpha) f(\beta) 
\left\{ (T \alpha^* \beta)^2 + T \alpha^* \beta \right\}
\left\langle \sqrt{T}\alpha \right| 
\left. \sqrt{T}\beta \right\rangle
\left\langle \sqrt{R}\alpha \right| 
\left. \sqrt{R}\beta \right\rangle
\nonumber\\ & = & 
T^2 \langle\chi|\hat{n}^2|\chi\rangle 
+ T(1-T) \langle\chi|\hat{n}|\chi\rangle ,
\end{eqnarray}
%%%%%%%%%%%%%%%%%%%%%%%%%%%
and 
%%%%%%%%%%%%%%%%%%%%%%%%%%%
\begin{eqnarray}
\langle\Phi| \hat{g}_1 |\Phi\rangle^2 & = & 
\left(
\int {\rm d}^2 \alpha \, \int {\rm d}^2 \beta \, f^*(\alpha) f(\beta) 
\left\langle \sqrt{T}\alpha \right|_A  
\left\langle \sqrt{R}\alpha \right|_B  
\left( 
\hat{a}^\dagger \hat{a} 
\right)
\left| \sqrt{T}\beta \right\rangle_A
\left| \sqrt{R}\beta \right\rangle_B
\right)^2 
\nonumber\\ & = & 
T^2 \langle\chi|\hat{n}|\chi\rangle^2. 
\end{eqnarray}
%%%%%%%%%%%%%%%%%%%%%%%%%%%
\end{widetext}
In total, we have 
%%%%%%%%%%%%%%%%%%%%%%%%%%%
\begin{eqnarray}
F_Q(|\chi\rangle,\hat{g}_1,T) & = & 4 (\langle\Phi| \hat{g}_1^2 |\Phi\rangle 
- \langle\Phi| \hat{g}_1 |\Phi\rangle^2) 
\nonumber\\ & = & 
4 \left\{ T^2 V_\chi + T(1-T) \bar{n}_\chi \right\}. 
\end{eqnarray}
%%%%%%%%%%%%%%%%%%%%%%%%%%%
For $T=1/2$, it is $V_\chi + \bar{n}_\chi$ and thus 
we get Eq.~(\ref{eq:qfi_g_1}).

\subsection{QFIM for $\hat{g}_d$ and $\hat{g}_s$ [Eq.~(\ref{eq:qfi_g_d}), (\ref{eq:QFIM_entry_dd})--(\ref{eq:QFIM_entry_ds})]}
For pure states, the elements of the QFIM are given by 
%%%%%%%%%%%%%%%%%%%%%%%%%%%
\begin{equation}
\label{eq:QFIM_entry}
F_{ij} = 4 \left( \langle \hat{g}_i \hat{g}_j \rangle 
- \langle \hat{g}_i \rangle \langle \hat{g}_j \rangle \right), 
\end{equation}
%%%%%%%%%%%%%%%%%%%%%%%%%%%
where $i,j$ takes $s$ and $d$.

Recall that 
$\hat{g}_d = (\hat{a}^\dagger \hat{a} - \hat{b}^\dagger \hat{b})/2$ and 
$\hat{g}_s = (\hat{a}^\dagger \hat{a} + \hat{b}^\dagger \hat{b})/2$. 
Then we have 
\begin{widetext}
%%%%%%%%%%%%%%%%%%%%%%%%%%%
\begin{eqnarray}
4 \langle\Phi| \hat{g}_d^2 |\Phi\rangle & = & 
\int {\rm d}^2 \alpha \, \int {\rm d}^2 \beta \, f^*(\alpha) f(\beta) 
\nonumber\\ && \times
\left\langle \sqrt{T}\alpha \right|_A  
\left\langle \sqrt{R}\alpha \right|_B  
\left( 
\hat{a}^{\dagger \, 2} \hat{a}^2 + \hat{a}^\dagger \hat{a} +
\hat{b}^{\dagger \, 2} \hat{b}^2 + \hat{b}^\dagger \hat{b} 
- 2 \hat{a}^\dagger \hat{a} \hat{b}^\dagger \hat{b} \right)
\left| \sqrt{T}\beta \right\rangle_A
\left| \sqrt{R}\beta \right\rangle_B
\nonumber\\ & = & 
\int {\rm d}^2 \alpha \, \int {\rm d}^2 \beta \, f^*(\alpha) f(\beta) 
\nonumber\\ && \times
\left\{ (T \alpha^* \beta)^2 + T \alpha^* \beta 
+ (R \alpha^* \beta)^2 + R \alpha^* \beta 
- 2 RT (\alpha^* \beta)^2 
\right\} 
\left\langle \sqrt{T}\alpha \right| 
\left. \sqrt{T}\beta \right\rangle
\left\langle \sqrt{R}\alpha \right| 
\left. \sqrt{R}\beta \right\rangle
\nonumber\\ & = & 
\int {\rm d}^2 \alpha \, \int {\rm d}^2 \beta \, f^*(\alpha) f(\beta) 
\left\{\alpha^* \beta + (T-R)^2 (\alpha^* \beta)^2 \right\} 
\exp\left[ -\frac{1}{2} \left( |\alpha|^2 + |\beta|^2 - 2 \alpha^* \beta 
\right) \right] 
\nonumber\\ & = & 
\langle\chi| \hat{n} |\chi\rangle 
+ (1-2T)^2 \left( \langle\chi| \hat{n}^2 |\chi\rangle 
- \langle\chi| \hat{n} |\chi\rangle \right). 
\end{eqnarray}
%%%%%%%%%%%%%%%%%%%%%%%%%%%
Similarly, we have 
%%%%%%%%%%%%%%%%%%%%%%%%%%%
\begin{eqnarray}
4 \langle\Phi| \hat{g}_s^2 |\Phi\rangle & = & 
\langle\chi| \hat{n}^2 |\chi\rangle ,
\\
4 \langle\Phi| \hat{g}_d \hat{g}_s |\Phi\rangle & = & 
4 \langle\Phi| \hat{g}_s \hat{g}_d |\Phi\rangle = 
-(1-2T) \langle\chi| \hat{n}^2 |\chi\rangle .
\end{eqnarray}
%%%%%%%%%%%%%%%%%%%%%%%%%%%
Also 
%%%%%%%%%%%%%%%%%%%%%%%%%%%
\begin{eqnarray}
2 \langle\Phi| \hat{g}_d |\Phi\rangle & = & 
\int {\rm d}^2 \alpha \, \int {\rm d}^2 \beta \, f^*(\alpha) f(\beta) 
\left\langle\sqrt{T}\alpha\right|_A 
\left\langle\sqrt{R}\alpha\right|_B 
\left( \hat{a}^\dagger \hat{a} - \hat{b}^\dagger \hat{b} \right) 
\left|\sqrt{T}\beta\right\rangle_A
\left|\sqrt{R}\beta\right\rangle_B 
\nonumber\\ & = & 
\int {\rm d}^2 \alpha \, \int {\rm d}^2 \beta \, f^*(\alpha) f(\beta) 
\left( T\alpha^* \beta - R\alpha^* \beta \right) 
\left\langle\sqrt{T}\alpha\right|
\left.\sqrt{T}\beta\right\rangle
\left\langle\sqrt{R}\alpha\right|
\left.\sqrt{R}\beta\right\rangle
\nonumber\\ & = & 
(1-2T) 
\int {\rm d}^2 \alpha \, \int {\rm d}^2 \beta \, f^*(\alpha) f(\beta) 
\alpha^* \beta 
\exp\left[ -\frac{1}{2} \left( |\alpha|^2 + |\beta|^2 - 2 \alpha^* \beta 
\right) \right] 
\nonumber\\ & = & 
(1-2T) \langle\chi| \hat{n} |\chi\rangle ,
\end{eqnarray}
%%%%%%%%%%%%%%%%%%%%%%%%%%%
\end{widetext}
and similarly, 
%%%%%%%%%%%%%%%%%%%%%%%%%%%
\begin{equation}
2 \langle\Phi| \hat{g}_s |\Phi\rangle = 
\langle\chi| \hat{n} |\chi\rangle .
\end{equation}
%%%%%%%%%%%%%%%%%%%%%%%%%%%
By using the above results, we have 
%%%%%%%%%%%%%%%%%%%%%%%%%%%
\begin{eqnarray}
\label{eq:qfi_3}
F_{dd} & = & F_Q(|\chi\rangle,\hat{g}_d,T) 
\nonumber\\ & = & 
\langle\chi| \hat{n} |\chi\rangle 
+ (1-2T)^2 \left( \langle\chi| \hat{n}^2 |\chi\rangle 
- \langle\chi| \hat{n} |\chi\rangle \right) 
\nonumber\\ && 
- (1-2T)^2 \langle\chi| \hat{n} |\chi\rangle^2 
\nonumber\\ & = & 
\left\{ 1- (1-2T)^2 \right\} \bar{n}_\chi 
+ (1-2T)^2 V_\chi ,
\end{eqnarray}
%%%%%%%%%%%%%%%%%%%%%%%%%%%
%%%%%%%%%%%%%%%%%%%%%%%%%%%
\begin{eqnarray}
F_{ss} & = & \langle\chi| \hat{n}^2 |\chi\rangle 
- \langle\chi| \hat{n} |\chi\rangle^2 
\nonumber\\ & = & 
V_\chi ,
\\
F_{ds} & = & F_{sd} = -(1-2T) \langle\chi| \hat{n}^2 |\chi\rangle 
-(1-2T) \langle\chi| \hat{n} |\chi\rangle^2 
\nonumber\\ & = & 
-(1-2T) V_\chi .
\end{eqnarray}
%%%%%%%%%%%%%%%%%%%%%%%%%%%

\section{Supplemental Material 2: Convexity of quantum Fisher 
information matrix}

Here we prove the convexity of the quantum Fisher information 
matrix (QFIM): 
%%%%%%%%%%%%%%%%%%%%%%%%%%%
\begin{equation}
\mathcal{F}_Q(\hat{\rho}_\varphi) \le p \mathcal{F}_Q(\hat{\sigma}_\varphi) 
+ (1-p) \mathcal{F}_Q(\hat{\tau}_\varphi),  
\end{equation}
%%%%%%%%%%%%%%%%%%%%%%%%%%%
for $\hat{\rho}_\varphi = p \hat{\sigma}_\varphi + (1-p) \hat{\tau}_\varphi$.
Here $\hat{\rho}_\varphi$, $\hat{\sigma}_\varphi$, and $\hat{\tau}_\varphi$ 
are (maybe mixed) quantum states where 
$\varphi = \{ \varphi_1 , \dots , \varphi_M \}$ 
is a set of $M$ unknown parameters. 

To begin with, we briefly review the definition and the structure of the QFIM 
that we will use in the proof. Detailed review on the QFI and QFIM 
can be found for example in Ref.~\cite{DJK15,HelstromBook,PG10}. 
The QFIM for $\hat{\rho}_\varphi$ is given by an $M \times M$ matrix 
$\mathcal{F}_Q(\hat{\rho}_\varphi) = [F_{ij}(\hat{\rho}_\varphi)]_{ij}$ 
($i,j = 1, \dots , M$) where each entry is defined as 
%%%%%%%%%%%%%%%%%%%%%%%%%%%
\begin{equation}
\label{eq:QFIM_def}
F_{ij}(\hat{\rho}_\varphi) = \frac{1}{2}
{\rm Tr}\left[ \hat{\rho}_\varphi \hat{L}_i \hat{L}_j + 
\hat{\rho}_\varphi \hat{L}_j \hat{L}_i \right] ,
\end{equation}
%%%%%%%%%%%%%%%%%%%%%%%%%%%
and $\hat{L}_i$, called the symmetrized logarithmic derivative, 
is a Hermitian operator satisfying 
%%%%%%%%%%%%%%%%%%%%%%%%%%%
\begin{equation}
\frac{\partial}{\partial \varphi_i} \hat{\rho}_\varphi = \frac{1}{2} 
\left( \hat{L}_i \hat{\rho}_\varphi + \hat{\rho}_\varphi \hat{L}_i
\right) .
\end{equation}
%%%%%%%%%%%%%%%%%%%%%%%%%%%
Let $\hat{\rho}_\varphi = \sum_k \lambda_k |\lambda_k\rangle\langle\lambda_k|$ 
be the spectral decomposition of $\hat{\rho}_\varphi$. 
Then we can explicitly describe $\hat{L}_i$ as 
%%%%%%%%%%%%%%%%%%%%%%%%%%%
\begin{equation}
\hat{L}_i = 2 \sum_{k,l} \frac{
\langle\lambda_k| \hat{\rho}_\varphi^{(i)} |\lambda_l\rangle }{
\lambda_k + \lambda_l}
|\lambda_k\rangle\langle\lambda_l| ,
\end{equation}
%%%%%%%%%%%%%%%%%%%%%%%%%%%
where $\hat{\rho}_\varphi^{(i)} = \partial \hat{\rho}_\varphi / 
\partial \varphi_i$.
Combining it with Eq.~(\ref{eq:QFIM_def}), the QFIM is expressed as 
%%%%%%%%%%%%%%%%%%%%%%%%%%%
\begin{equation}
\label{eq:QFIM_def2}
F_{ij}(\hat{\rho}_\varphi) = 2 \sum_{k,l} 
\frac{
 \langle\lambda_k| \hat{\rho}_\varphi^{(i)} |\lambda_l\rangle 
 \langle\lambda_l| \hat{\rho}_\varphi^{(j)} |\lambda_k\rangle 
}{
\lambda_k + \lambda_l} .
\end{equation}
%%%%%%%%%%%%%%%%%%%%%%%%%%% 
We also use an important property of the QFIM: 
monotonicity under completely positive trace preserving (CPTP) map 
$\mathcal{L}$ \cite{P96,PG10},  
%%%%%%%%%%%%%%%%%%%%%%%%%%%
\begin{equation}
\label{eq:QFIM_monotonicity}
\mathcal{F}_Q (\hat{\rho}_\varphi) 
\ge \mathcal{F}_Q (\mathcal{L}(\hat{\rho}_\varphi)). 
\end{equation}
%%%%%%%%%%%%%%%%%%%%%%%%%%%

The proof of the convexity of the QFIM is basically given by extending 
the proof for the QFI (i.e. single-parameter case) in Ref.~\cite{F01}. 
Consider the bipartite state $\tilde{\rho}_\varphi^{AB} 
= p |e_0 \rangle\langle e_0|^A \otimes \hat{\sigma}_\varphi^B 
+ (1-p) |e_1 \rangle\langle e_1|^A \otimes \hat{\tau}_\varphi^B$, 
where $|e_k\rangle$ is an orthonormal basis in $A$. 
Note that ${\rm Tr}_A [\tilde{\rho}_\varphi^{AB}] = \hat{\rho}_\varphi^B$. 
Then we have 
%%%%%%%%%%%%%%%%%%%%%%%%%%%
\begin{equation}
\label{eq:rho_tilde}
\mathcal{F}_Q(\tilde{\rho}_\varphi^{AB}) = 
p \mathcal{F}_Q (\hat{\sigma}_\varphi^B) 
+ (1-p) \mathcal{F}_Q (\hat{\tau}_\varphi^B) .
\end{equation}
%%%%%%%%%%%%%%%%%%%%%%%%%%%
This is justified by the following observation. 
Since $|e_k\rangle$ is independent of the unknown parameters $\varphi_i$, 
$\tilde{\rho}_\varphi^{(i)} = 
p |e_0 \rangle\langle e_0| \otimes \hat{\sigma}_\varphi^{(i)} 
+ (1-p) |e_1 \rangle\langle e_1| \otimes \hat{\tau}_\varphi^{(i)}$, 
for any $i$. 
Also the spectral decomposition of $\tilde{\rho}_\varphi$ 
is described as 
$p |e_0 \rangle\langle e_0| \otimes \sum_i \lambda_i^{\sigma} 
|\lambda_i^{\sigma}\rangle\langle\lambda_i^{\sigma}| + 
(1-p) |e_1 \rangle\langle e_1| \otimes \sum_i \lambda_i^{\tau} 
|\lambda_i^{\tau}\rangle\langle\lambda_i^{\tau}|$, 
where $\sum_i \lambda_i^{\sigma} 
|\lambda_i^{\sigma}\rangle\langle\lambda_i^{\sigma}|$ and 
$\sum_i \lambda_i^{\tau} 
|\lambda_i^{\tau}\rangle\langle\lambda_i^{\tau}|$ are 
the spectral decompositions of $\hat{\sigma}_\varphi$ and 
$\hat{\tau}_\varphi$, respectively. 
Plugging them into the expression of QFI in Eq.~(\ref{eq:QFIM_def2}), 
we get 
%%%%%%%%%%%%%%%%%%%%%%%%%%%
\begin{equation}
F_{ij}(\tilde{\rho}_\varphi^{AB}) = 
p F_{ij}(\hat{\sigma}_\varphi^B) + (1-p) F_{ij}(\hat{\tau}_\varphi^B). 
\end{equation}
%%%%%%%%%%%%%%%%%%%%%%%%%%%
Since this holds for all $i$ and $j$, we get Eq.~(\ref{eq:rho_tilde}).

By using Eq.~(\ref{eq:rho_tilde}), 
the monotonicity (\ref{eq:QFIM_monotonicity}), and 
the fact that partial trace is a CPTP map, we have 
%%%%%%%%%%%%%%%%%%%%%%%%%%%
\begin{eqnarray}
\mathcal{F}_Q(\hat{\rho}_\varphi^B) & \le & 
\mathcal{F}_Q(\hat{\rho}_\varphi^{AB}) 
\nonumber\\ & = & 
p \mathcal{F}_Q (\hat{\sigma}_\varphi^B) 
+ (1-p) \mathcal{F}_Q (\hat{\tau}_\varphi^B) ,
\end{eqnarray}
%%%%%%%%%%%%%%%%%%%%%%%%%%%
which completes the proof of the convexity of the QFIM.

\section{Supplemental Material 3: Quantum Fisher information 
for $\hat{g}_1$ with phase randomizing}

Here we calculate the QFI for $|n\rangle \otimes |0\rangle$ with 
the generator $\hat{g}_1 = \hat{a}^\dagger \hat{a}$ and 
see that it coincides with that of $\hat{g}_d$. 
The state past the beam splitter and the phase-shift transformation 
is given by
\begin{eqnarray}
|\psi_n(\phi)\rangle _{AB} & = & 
\sum_{j=0}^{n}e^{-ij\phi}{n \choose j}^{1/2}
\nonumber\\ && \times 
T^{j/2}(1-T)^{(n-j)/2}|j\rangle _A\otimes|n-j\rangle_B. \nonumber\\
\end{eqnarray}
For $|\psi_n(\phi)\rangle$, we find 
\begin{align}
\langle \hat{a}^{\dagger}\hat{a}\rangle  & = 
\sum_{j=0}^{n}j{n \choose j}T^{j}\left(1-T\right)^{n-j}=nT,\\
\langle \hat{b}^{\dagger}\hat{b}\rangle  & = n\left(1-T\right),\\
\langle \hat{a}^{\dagger2}\hat{a}^{2}\rangle &=  
\sum_{j=0}^{n}j\left(j-1\right){n \choose j}T^{j}\left(1-T\right)^{n-j} 
\nonumber\\ & = n(n-1)T^2 ,
\end{align}
and the QFI evaluated as $4\left(\langle \hat{g}_1^2\rangle -\langle \hat{g}_{1}\rangle 
^{2}\right)$ is found to be
\begin{align}
F_{Q}^{\left(1\right)} & =4\left(\langle \hat{a}^{\dagger2}\hat{a}^{2}\rangle 
+\langle \hat{a}^{\dagger}\hat{a}\rangle -\langle \hat{a}^{\dagger}\hat{a}\rangle 
^{2}\right)\nonumber\\
 & =4\left\{ n(n-1)T^2 + nT - n^2 T^2 \right\}\nonumber\\
 & =4nT\left(1-T\right).
\end{align}
The maximum is attained at $T=1/2$ and is equal to $n$.

\section{Supplemental Material 4: Derivation of $F(\phi_A)$
}

The calculation of Fisher information can be performed by 
the characteristic function approach. 
For the details of the characteristic function formalism in quantum optics, 
see Ref.~\cite{WPGCRSL12} for example. 
Here we follow the definition and the methodology developed 
in Ref.~\cite{TJS15}. 
Then the covariance matrix of the two-mode vacuum is given by 
%%%%%%%%%%%%%%%%%%%%%%%%%%%
\begin{equation}
\label{eq:vacuum input}
\gamma_{\rm in} = I(4), 
\end{equation}
%%%%%%%%%%%%%%%%%%%%%%%%%%%
where $I(4)$ is the four-by-four identity matrix. 
The beam splitter unitary transformation is represented by 
the symplectic transformation: 
%%%%%%%%%%%%%%%%%%%%%%%%%%%
\begin{equation}
\label{eq:S_BS}
S_{\rm BS} = \left[
\begin{array}{cccc}
\sqrt{T} & 0 &  \sqrt{1-T} & 0 \\
0 & \sqrt{T} & 0 & \sqrt{1-T} \\
-\sqrt{1-T} & 0 & \sqrt{T} & 0 \\
0 & -\sqrt{1-T} & 0 & \sqrt{T} 
\end{array}
\right]
\end{equation}
%%%%%%%%%%%%%%%%%%%%%%%%%%%
Similarly, the unknown phase shift is given by 
%%%%%%%%%%%%%%%%%%%%%%%%%%%
\begin{equation}
\label{eq:S_PS}
S_{\rm PS} = \left[
\begin{array}{cccc}
\cos\phi_A & \sin\phi_A & 0 & 0 \\
-\sin\phi_A & \cos\phi_A & 0 & 0 \\
0 & 0 & 1 & 0 \\
0 & 0 & 0 & 1 
\end{array}
\right], 
\end{equation}
%%%%%%%%%%%%%%%%%%%%%%%%%%%
and the squeezing in the first arm is given by 
%%%%%%%%%%%%%%%%%%%%%%%%%%%
\begin{equation}
\label{eq:S_BS}
S_{\rm SQ}(r) = \left[
\begin{array}{cccc}
e^{-r} & 0 &  0 & 0 \\
0 & e^{r} & 0 & 0 \\
0 & 0 & 1 & 0 \\
0 & 0 & 0 & 1 
\end{array}
\right] ,
\end{equation}
%%%%%%%%%%%%%%%%%%%%%%%%%%%
where $e^{-r} = \sqrt{\bar{n}+1} - \sqrt{\bar{n}}$ 
and $e^r = \sqrt{\bar{n}+1} + \sqrt{\bar{n}}$
(remember $\bar{n} = \sinh^2 r$). 

Then the covariance matrix of the state before 
the photo detectors is calculated to be

%%%%%%%%%%%%%%%%%%%%%%%%%%%
\begin{eqnarray}
\label{eq:gamma_out}
\gamma_{\rm out} & = & 
S_{SQ}(-r) S_{\rm BS}^T S_{\rm PS} S_{\rm BS} S_{SQ}(r) 
\gamma_{\rm in} 
\nonumber\\ && \times
S_{SQ}^T(r) S_{\rm BS}^T S_{\rm PS}^T S_{\rm BS} S_{SQ}^T (-r)
\end{eqnarray}
%%%%%%%%%%%%%%%%%%%%%%%%%%%
where the superscript $T$ denotes the matrix transpose. 

The probability of having no-clicks at both detector 
(i.e. the projection onto $|0\rangle\langle0|\otimes|0\rangle\langle0|$) 
is given by \cite{TJS15},
%%%%%%%%%%%%%%%%%%%%%%%%%%%
\begin{equation}
\label{eq:P00}
P_{00} = \frac{4}{\sqrt{\det (\gamma_{\rm out} + I(4))}} .
\end{equation}
%%%%%%%%%%%%%%%%%%%%%%%%%%%

Then the Fisher information for $\phi_A$ is calculated by 
%%%%%%%%%%%%%%%%%%%%%%%%%%%
\begin{equation}
\label{eq:CFI}
F(\phi_1) = \frac{1}{P_{00}} \left(\frac{d P_{00}}{d \phi_A}\right)^2 
+ \frac{1}{1-P_{00}} \left(\frac{d (1-P_{00})}{d \phi_A}\right)^2 .
\end{equation}
%%%%%%%%%%%%%%%%%%%%%%%%%%%
The calculation is performed by Mathematica. 
Since the expression of $F(\phi_A)$ is quite complicated, 
we consider the limit of small $\phi_A$. 
Then we get 
%%%%%%%%%%%%%%%%%%%%%%%%%%%
\begin{equation}
\label{eq:CFI}
\lim_{\phi_A \to 0} F = 
4 \bar{n} T (1+ T + 2 \bar{n}T) .
\end{equation}
%%%%%%%%%%%%%%%%%%%%%%%%%%%
Replacing $\bar{n}$ with $\bar{n}_{\rm tot} = 2\bar{n}$, we get 
%%%%%%%%%%%%%%%%%%%%%%%%%%%
\begin{equation}
\label{eq:CFI}
\lim_{\phi_A \to 0} F = 
2 \bar{n}_{\rm tot} T (1+ T + \bar{n}_{\rm tot} T) ,
\end{equation}
%%%%%%%%%%%%%%%%%%%%%%%%%%%
which implies that in the limit of small phase shifts, 
the Fisher information of our protocol can surpass the SNL in terms of 
the total resource for any $T \ne 0$, and particularly for $T=1/2$. 

\end{document}